\documentstyle[graphicx,multicol,eqsecnum,prb,aps]{revtex}

\begin{document}
\draft
\preprint{HEP/123-qed}
\title {Far-Infrared Spectroscopy in Spin-Peierls Compound CuGeO$_{3}$ under High Magnetic Fields}
\author {K. Takehana, T. Takamasu, M. Hase, G. Kido}
\address {Physical Properties Division, National Research Institute for Metals,\\
3-13 Sakura, Tsukuba-city, Ibaraki 305-0003, Japan}
\author{K. Uchinokura}
\address {Department of Advanced Materials Science, The Univ. of Tokyo,\\ 6th Engineering Bld., 7-3-1 Hongo, Bunkyo-ku, Tokyo 113-8656, Japan}
\date{\today}
\begin{multicols}{2}[
\maketitle
\begin{abstract}
Polarized far-infrared (FIR) spectroscopic measurements and FIR magneto-optical studies were performed on the inorganic spin-Peierls compound CuGeO$_{3}$. An absorption line, which was found at 98 cm$^{-1}$ in the dimerized phase (D phase), was assigned to a folded phonon mode of B$_{3u}$ symmetry. The splitting of the folded mode into two components in the incommensurate phase (IC phase) has been observed for the first time.   A new broad absorption centered at 63 cm$^{-1}$ was observed only in the ${\bf E}\parallel b$ axis polarization, which was assigned to a magnetic excitation from singlet ground state to a continuum state.
\end{abstract}
]
\narrowtext

\section{Introduction}
\label{sec:level1}

The discovery of the spin-Peierls (SP) transition in an inorganic 
compound CuGeO$_{3}$ (Ref.~\onlinecite{Hase}) has produced great interest in the 
properties of this quasi-one-dimensional $S = 1/2$ Heisenberg 
antiferromagnet. The lattice dimerization induced by the SP transition 
was confirmed by observation of superlattice Bragg peaks by electron 
diffraction,\cite{Kaminura} x-ray and neutron diffraction.\cite{Pouget,Hitota} At low temperatures, a high magnetic field above 12 T induces a phase 
transition from the dimerized phase (D phase) to a magnetic phase.\cite{Hase2} The high-field phase was assigned to an incommensurate phase (IC 
phase) based on the appearance of a splitting of the superlattice 
reflections above the critical field, $H_C$.\cite{Kiryukhin1} Higher order harmonics of the incommensurate Bragg reflections, which indicates that the lattice modulation forms a soliton lattice, were 
observed just above $H_C$ by the x-ray experiments.\cite{Kiryukhin2} An anomalous spontaneous strain appears with the SP transition and increases with 
decreasing temperature.\cite{Lorenzo} This strain is partially removed at the 
transition from the D to the IC phase and continues to decrease 
with increasing field, as observed by magnetostriction measurements 
under high magnetic fields.\cite{Takehana1,Takehana2}

Since the coupling between the one-dimensional spin system and three-dimensional phonon fields plays an essential role in the SP transition, it is 
important to investigate the phonons related with the transition such as folded phonons and a soft phonon. The crystal structure of CuGeO$_{3}$ belongs to 
Pbmm symmetry at room temperature.\cite{Vollenkle} According to the factor group analysis, optically active phonons are 12 modes (4A$_{g}$($aa, bb, 
cc$) + 4B$_{1g}$($ab$) + 3B$_{2g}$($ac$) + B$_{3g}$($bc$)) in the Raman spectra and 13 modes (3B$_{1u}$(${\bf E}\parallel c$) + 5B$_{2u}$(${\bf E}\parallel b$) +5B$_{3u}$(${\bf E}\parallel a$)) in the infrared spectra. All of them have 
already been assigned in the Raman and infrared spectra.\cite{Popovic} In the SP state, the symmetry is lowered to Bbcm due to the formation of the 
superlattice\cite{Hitota,Braden} and additional 18 Raman active modes  (4A$_{g}$($aa, bb, cc$) + 5B$_{1g}$($ab$) + 4B$_{2g}$($ac$) + 5B$_{3g}$($bc$)) 
and 9 infrared active modes (2B$_{1u}$(${\bf E}\parallel c$) + 4B$_{2u}$(${\bf E}\parallel b$) + 3B$_{3u}$(${\bf E}\parallel a$)) should appear below $T_{SP}$.  
At present, three Raman modes at 107, 369 and 820  cm$^{-1}$ were assigned to the A$_{g}$ folded phonons in the D phase, while the first one has a Fano type line shape.\cite{Ogita,Loa,Loosdrecht} 
Concerning the infrared active folded phonons, one B$_{1u}$ and two B$_{2u}$ modes were found at 284.2, 311.7 and 800 cm$^{-1}$, respectively.\cite{Damascelli,Popova} 
Besides those 3 infrared active modes, we assigned the absorption line at 98 cm$^{-1}$ to a folded phonon in our previous paper, while its polarization properties were unclear.\cite{Takehana3}

The field dependence of the folded phonon modes was investigated in the Raman experiments.\cite{Loosdrecht2,Loa2} The intensity decreases steeply at the boundary between the D and IC phases, while no energy shift was observed. 
Just above $H_C$, the intensity decreases to about half of the D phase, and continues to decrease in the IC phase with increasing field.\cite{Loa2} 
In contrast to these results, however, the folded phonon mode at 98 cm$^{-1}$ was unobservable in the IC phase in our previous paper.\cite{Takehana3}

Magnetic excitations across the gap between the singlet ground state and the triplet excited state were found below $T_{SP}$ in the far-infrared 
(FIR) measurements\cite{Takehana3,Brill,Li,Loosdrecht3,Nojiri} and continuous excitations between 30 and 230 cm$^{-1}$ due to two magnon processes were observed below 60 K in the Raman experiments.\cite{Ogita,Loa,Loosdrecht,Kuroe} 
Interestingly, two magnetic excitations were found at 19 cm$^{-1}$ and 44 cm$^{-1}$ by the electron spin resonance (ESR) measurements.\cite{Brill,Nojiri} 
Both modes split into two components in the presence of the magnetic field and become unobservable in the IC phase.   In a recent inelastic neutron scattering (INS) experiment, Lorenzo {\em et al}. found a second magnetic excitation branch with a non-negligible spectral weight, which indicates that there are two magnetic excitation gaps of 19 cm$^{-1}$ and 44 cm$^{-1}$ at the zone-center.\cite{Lorenzo2} 
The authors concluded that its existence originates from the presence of the Dzyaloshinskii-Moriya (DM) antisymmetric exchange terms, whose existence has also been suggested from ESR measurements.\cite{Yamada} 
In principle, magnetic excitations between the singlet and the triplet states is infrared inactive. Uhrig suggested that the observation of the gap at 44 cm$^{-1}$ is caused by the presence of the staggered magnetic field in analogy to NENP and NINO.\cite{Uhrig} 
Their model does not explain the observation in zero field and the field dependence of the intensity, so the microscopic mechanism of the magnetic excitation at 44 cm$^{-1}$ is still unclear. Three new branches were found in the IC phase by the FIR experiments.\cite{Loosdrecht3} 
One of them originates from the discommensuration because of its appearance only just above $H_C$ and others were explained as the absorption involving magnetic and vibrational excitations in a modulated structure. 
On the other hand, in the IC phase, suppression of the continuous structure between 30 and 230 cm$^{-1}$, especially the peak at 30 cm$^{-1}$,\cite{Loosdrecht2} and enhancement of the peak at 17 cm$^{-1}$ were observed in the Raman experiments.\cite{Loa2}

In order to clarify the property of the 98 cm$^{-1}$ mode, we presented in this paper the polarized spectroscopic measurements on CuGeO$_{3}$ in zero field and magneto-optical studies up to 18 T.

\section{Experimental}

A CuGeO$_{3}$ single crystal was grown by a floating zone method using an image furnace and was cleaved along the (100) plane.  
We used a wedge-shaped sample with dimension of 1.5 $\times$ 4 $\times$ 6 mm$^{3}$ in order to avoid the interference of FIR light in the sample.

FIR transmission was measured in the spectral range between 15 and 300 cm$^{-1}$ with a minimum resolution of 0.1 cm$^{-1}$ using a Fourier transform spectrometer (BOMEM DA8). 
The spectral range was covered by 50, 100 $\mu$m Mylar and Ge, Si coated 6 $\mu$m Mylar beamsplitters. An FIR polariser of free-standing wire grid type with a grid period of 12.5 $\mu$m was used for polarized measurements. 
A Si-bolometer which was operated at 4.2 K and a Si composite bolometer operated at 0.3 K were employed as the FIR detectors. 
The former was used for the higher energy region than 50 cm$^{-1}$ and the latter for the lower region. A high pressure mercury lamp was used as a light source. 
The unpolarized spectra in the presence of magnetic field were obtained with an 18 T superconducting magnet in the Faraday configuration. Both the spectrometer and the detector were placed 3 meters apart from the magnet in order to avoid the leakage flux of the magnet. 
The FIR light was led by an optical-pipe system, which consisted of well-polished brass pipes (i.d. 10 mm) and mirror cells which turn the FIR light direction through a right angle.\cite{Takehana4} 
All the optical path is evacuated to avoid the absorption due to water vapor. The temperature dependence of the spectra was investigated down to 3 K.

\section{Results}
\subsection{{\em B} = 0 T}

\begin{figure}
\centerline{\includegraphics*[width=8cm]{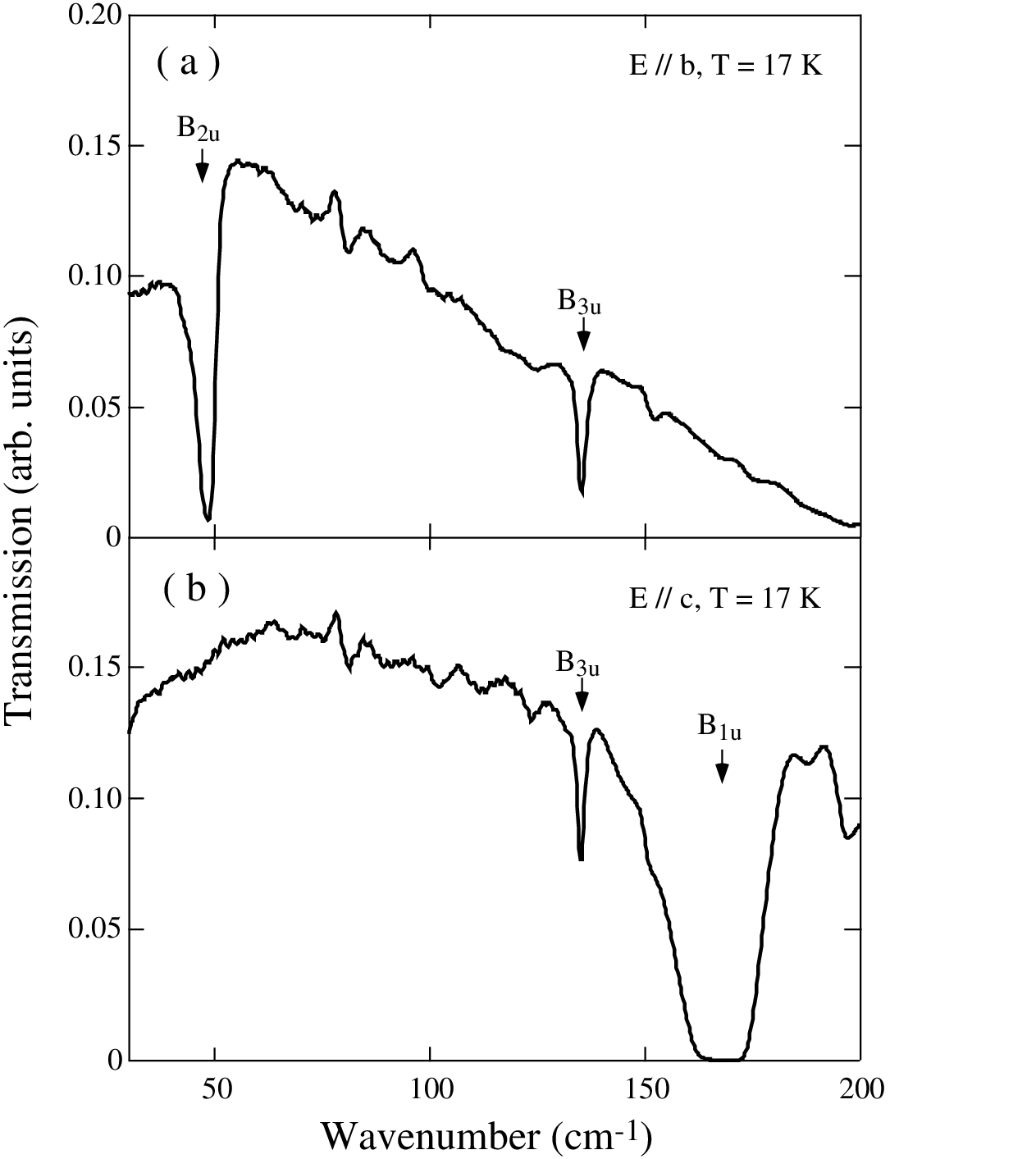}}
\caption{ Polarized transmission spectra at {\em B} = 0 T and {\em T} = 17 K in the (a) ${\bf E}\parallel b$ axis and (b) ${\bf E}\parallel c$ axis configurations.}
\label{fig1}
\end{figure}

Figures~\ref{fig1} (a) and (b) show the transmission spectrum of CuGeO$_{3}$ in the ${\bf E}\parallel b$ and ${\bf E}\parallel c$ axis configuration, respectively, at  {\em T} = 17 K. 
Absorptions due to the optical phonons are clearly observed at 48 cm$^{-1}$ and 135 cm$^{-1}$ in Fig.~\ref{fig1}(a) and at 135 and 170 cm$^{-1}$ in Fig.~\ref{fig1}(b), which were assigned to B$_{2u}$, B$_{3u}$, B$_{3u}$ and B$_{1u}$ modes, respectively, by the infrared reflectivity measurements.\cite{Popovic} 
The appearance of B$_{3u}$ modes in both the ${\bf E}\parallel b$ and ${\bf E}\parallel c$ configurations would be due to an inclination of a few degrees from the correct configuration.

\begin{figure}
\centerline{\includegraphics*[width=8cm]{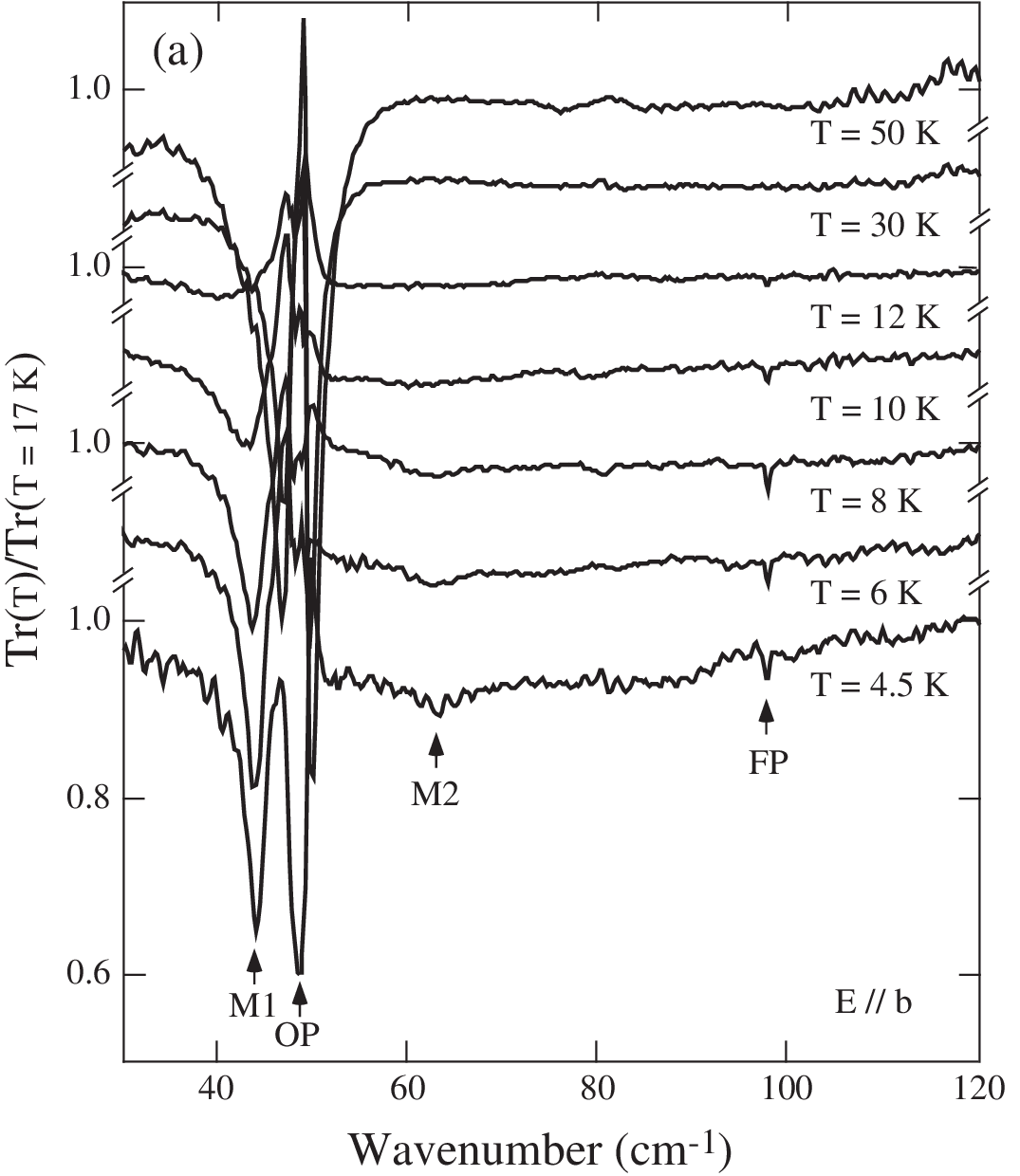}}
\centerline{\includegraphics*[width=8cm]{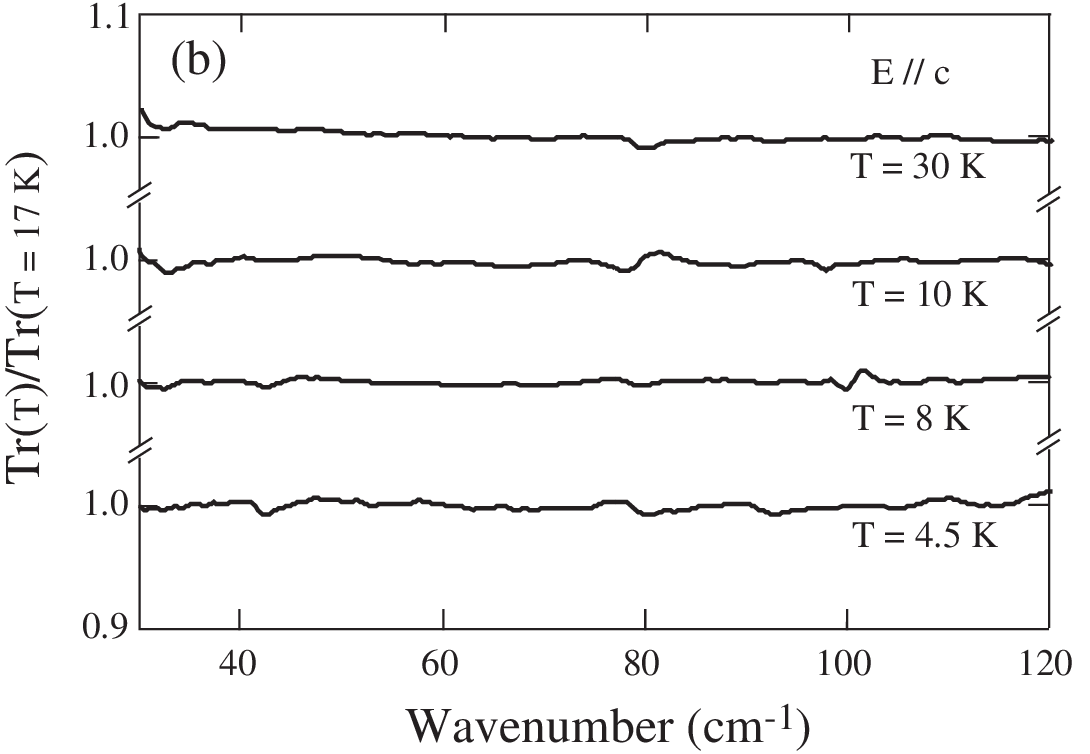}}
\caption{(a)  Normalized spectra, Tr({\em T})/Tr({\em T} = 17 K), in the configuration of the ${\bf E}\parallel b$ axis polarization at {\em B} = 0 T. The absorptions at 44 cm$^{-1}$, 98 cm$^{-1}$ and 63 cm$^{-1}$ below $T_{SP}$ grow with decreasing temperature. The structures around 50 cm$^{-1}$, which appear in all spectra, are caused by temperature dependence of the B$_{2u}$ optical phonon at 48 cm$^{-1}$.
(b)  Normalized spectra, Tr({\em T})/Tr({\em T} = 17 K), in the configuration of the ${\bf E}\parallel c$ axis polarization at {\em B} = 0 T. No significant structure was observed.}
\label{fig2}
\end{figure}

In order to clarify the small change between the three phases; i.e., the uniform (U), D and IC phase, the transmission spectra were normalized by the spectrum in the U phase, (Tr({\em T} = 17 K, {\em B} = 0 T)). 
Figures~\ref{fig2} (a) and (b) show the normalized spectra in the ${\bf E}\parallel b$ and ${\bf E}\parallel c$ configurations, respectively. A sharp absorption line at 98 cm$^{-1}$, which is labeled as FP in Fig.~\ref{fig2}(a), appears only in the ${\bf E}\parallel b$ configuration below $T_{SP}$. 
The absorption intensity of FP decreases without broadening when the temperature increases up to $T_{SP}$ and it shows no energy shift as a function of temperature. No structure was observed around 98 cm$^{-1}$ in the 30 K spectrum, which indicates that FP appears only below $T_{SP}$. 
An asymmetric absorption, M1, which has a tail on the lower-energy side, was observed at 44 cm$^{-1}$ only in the spectra of ${\bf E}\parallel b$ (${\bf H}\parallel c$) configuration.\cite{Damascelli} M1 loses its intensity rapidly with broadening and its peak position shifts slightly toward lower energy when the temperature increases up to $T_{SP}$. 
The broad absorption, M2, which is centered at 63 cm$^{-1}$ and has a wide tail on the higher-energy side, was newly found only in the ${\bf E}\parallel b$ (${\bf H}\parallel c$) configuration below $T_{SP}$ and grows with decreasing temperature, whose temperature dependence and polarization property are quite similar to that of M1. 
The structures around 50 cm$^{-1}$ in the ${\bf E}\parallel b$ configuration, OP, are due to the temperature dependence of the B$_{2u}$ optical phonon.  On the other hand, no significant structure was observed in the spectra of ${\bf E}\parallel c$ configuration in both the U and the D phases.

\begin{figure}
\centerline{\includegraphics*[width=8cm]{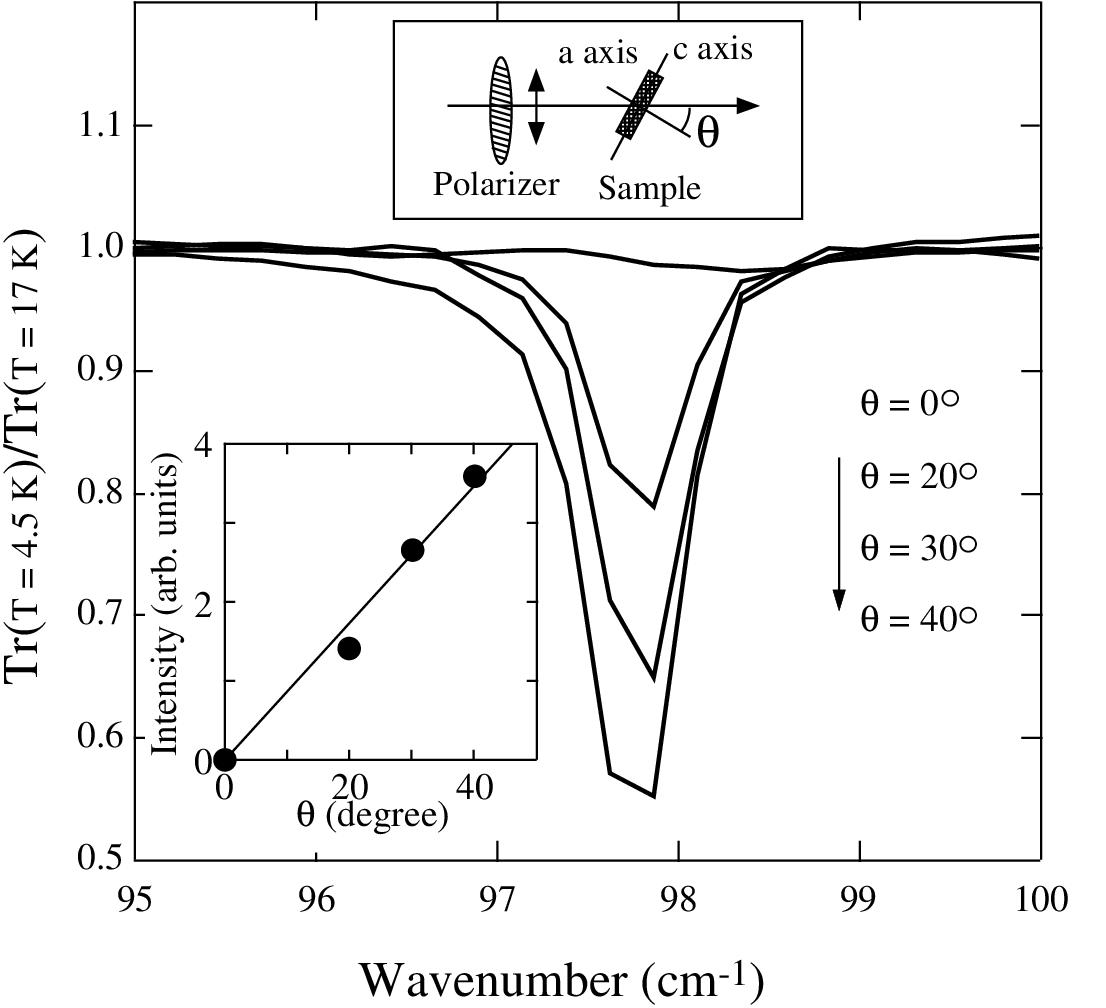}}
\caption{Angular dependence of the absorption at 98 cm$^{-1}$ for rotation of the sample around the $b$ axis (see upper inset). As shown in the lower inset, the intensity increases almost linearly  up to 40$^{\circ}$.}
\label{fig3}
\end{figure}

\begin{figure}
\centerline{\includegraphics*[width=8cm]{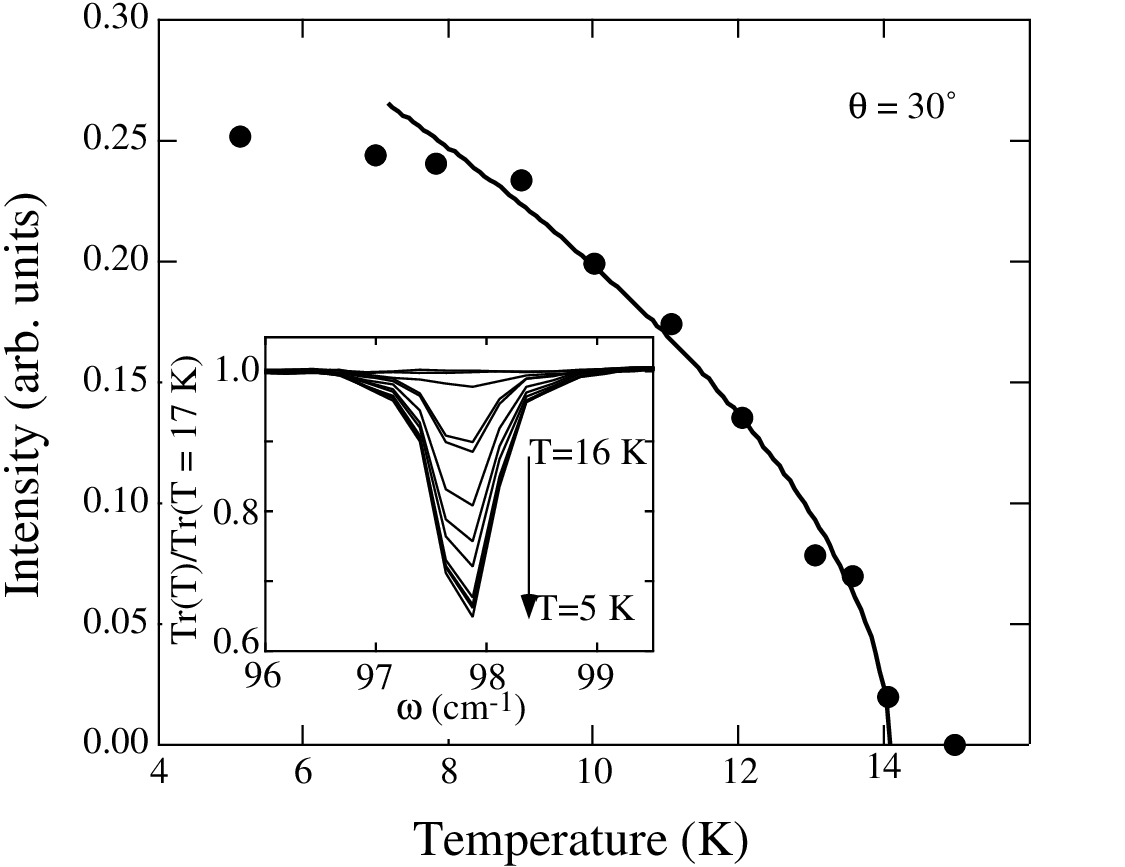}}
\caption{Temperature dependence of the intensity of the absorption at 98 cm$^{-1}$, when the sample is rotated by 30$^{\circ}$ around $b$ axis, which is well described by the power law, \(\alpha(T_{SP}-T)^{2\beta}\). The inset shows that the absorption at 98 cm$^{-1}$ increases with decreasing temperature.}
\label{fig4}
\end{figure}

In order to examine the polarization property of FP mode, an angular dependence was measured as shown in Fig.~\ref{fig3}, when the sample is rotated around the $b$ axis (see upper inset of Fig.~\ref{fig3}), which means that the component of ${\bf E}\parallel a$ polarization becomes mixed into the spectrum when the angle $\theta$ increases from zero degrees. 
The absorption intensity increases almost linearly without energy shift and broadening, as $\theta$ increases up to 40$^{\circ}$ (see lower inset of Fig.~\ref{fig3}). According to these results, we conclude that the FP mode must have the ${\bf E}\parallel a$ polarization property. 
Observation of this mode in the ${\bf E}\parallel b$ configuration, as shown in Fig.~\ref{fig2}(a), would be caused by the sample tilting by a few degrees from the correct configuration, because the modes of ${\bf E}\parallel a$ polarization must not be observed in the correct ${\bf E}\parallel b$ configuration. 
The absorption intensity at $\theta$ = 30$^{\circ}$ increases with decreasing temperature, as shown in the inset of Fig.~\ref{fig4}. The temperature dependence of its intensity is well described by the power law \(\alpha(T_{\rm SP}-T)^{2\beta}\) (see Fig.~\ref{fig4}). 
The best fit was obtained for 2$\beta$ = 0.55, which is in good agreement with that of other folded phonons,\cite{Damascelli,Popova} the superlattice reflections\cite{Hitota} and the spontaneous strains.\cite{Harris} 
The intensity of folded phonons in zero field is proportional to the square of the lattice distortion induced by the SP transition, as well as that of superlattice reflections and the spontaneous strains.\cite{Harris}

\subsection{{\em B} $\neq$ 0 T}

\begin{figure}
\centerline{\includegraphics*[width=8cm]{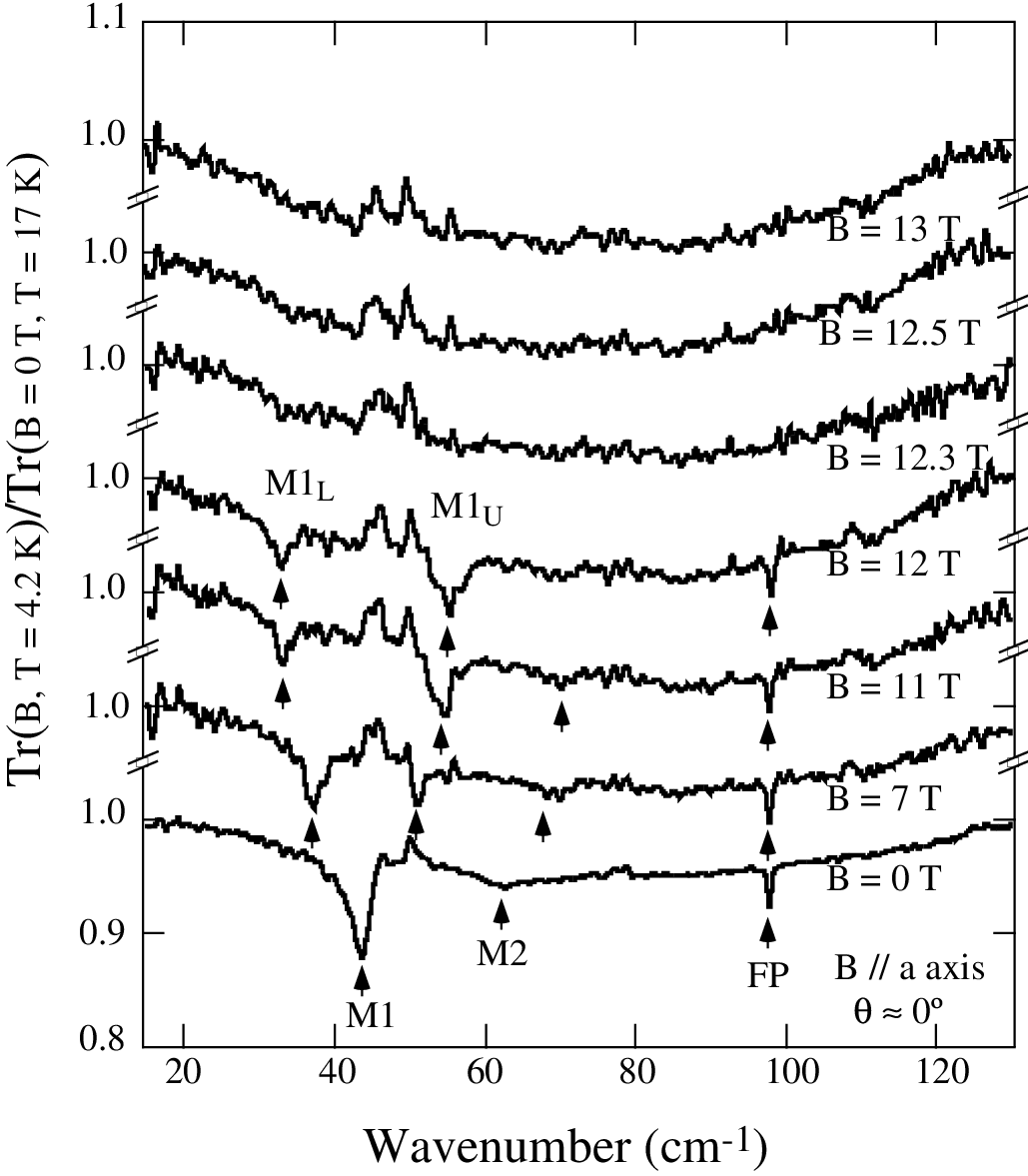}}
\caption{Unpolarized normalized spectra, Tr({\em B}, {\em T} = 4.2 K)/Tr({\em B} = 0 T, {\em T} = 17 K), for various fields, when the sample was tilted by a few degrees from the ${\bf B}\parallel a$ axis configuration.}
\label{fig5}
\end{figure}

The normalized spectra in the presence of magnetic fields are shown in Fig.~\ref{fig5}. The FP mode was observed in this configuration, which may be caused by an inclination of the sample, as mentioned above. Both peak position and peak intensity remain unchanged up to $H_C$ and become unclear above $H_C$. 
In the presence of a magnetic field, M1 splits into two components, M1$_{\rm U}$ and M1$_{\rm L}$, which correspond to the excitations from the singlet ground state to the $m_S = \pm 1$ branches of the triplet excited states. 
From the field dependence of the peak positions, we estimate the $g$-value as $g = 2.1$, which is in good agreement with the reported values.\cite{Brill,Loosdrecht3} Neither M1$_{\rm U}$ nor M1$_{\rm L}$ are observed in the IC phase, which is again in agreement with the previous results.\cite{Brill,Li,Loosdrecht3,Nojiri} 
With increase of magnetic field M2 shifts toward higher energy, while it could not be confirmed whether the lower branch exists or not, owing to overlap with M1$_{\rm U}$ branch. Around and above $H_C$, M2 becomes broader. There is no significant change of M2 at $H_C$.

\begin{figure}
\centerline{\includegraphics*[width=8cm]{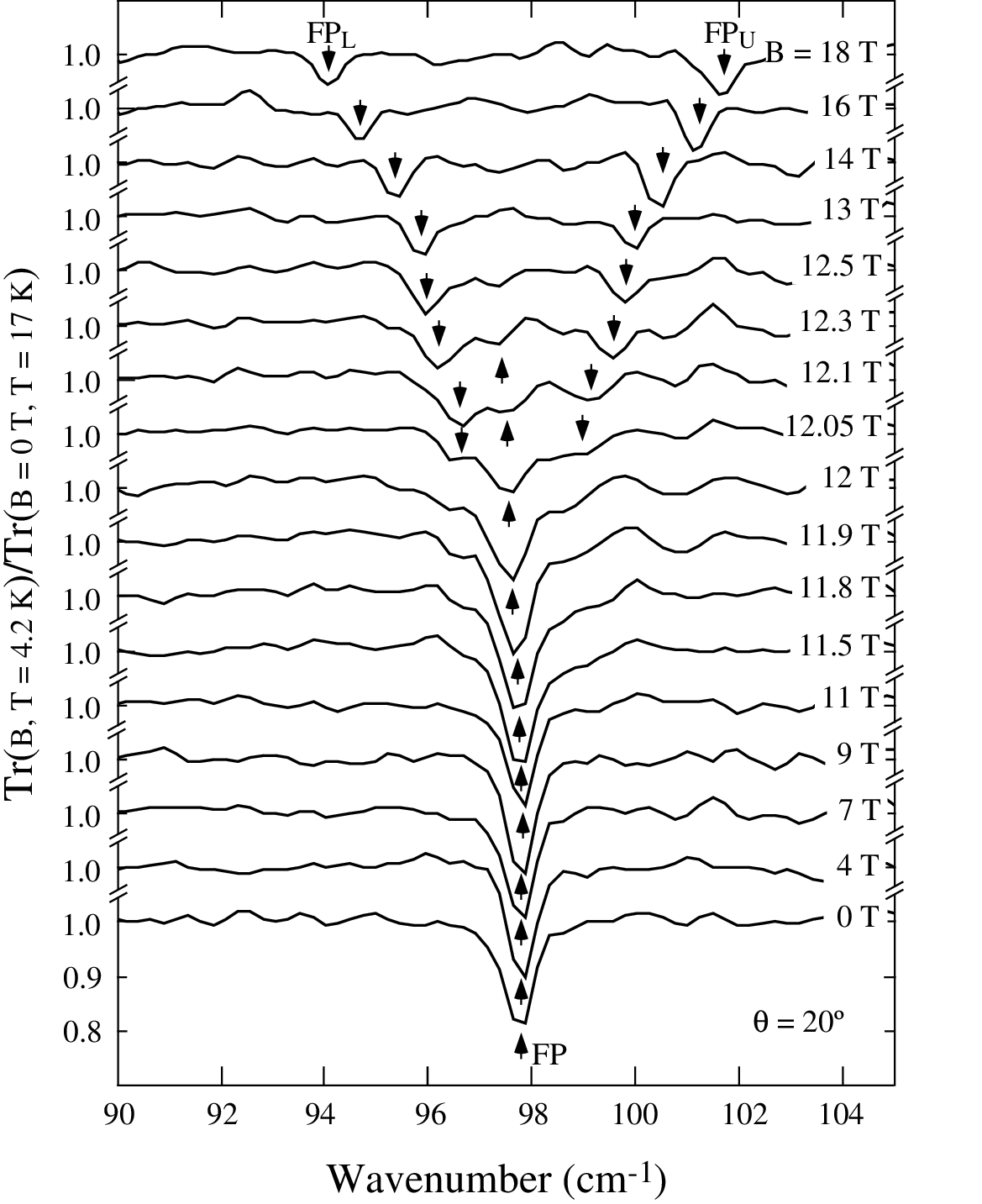}}
\caption{Field dependence of the absorption at 98 cm$^{-1}$ at 4.2 K, when the sample is rotated by 20$^{\circ}$. Satellite peaks appear on both sides of the absorption at 98 cm$^{-1}$ in the IC phase, while the main peak weakens and disappears around $H_C$.}
\label{fig6}
\end{figure}

In order to clarify the behavior of FP mode in the IC phase, the field dependence of the spectrum was investigated up to 18 T with the sample inclined around the $b$ axis by about 20$^{\circ}$ (see Fig.~\ref{fig6}). 
We have observed that two absorption lines, FP$_{\rm U}$ and FP$_{\rm L}$, appear on both sides of FP with approximately the same intervals around $H_C$, while FP loses its intensity with slightly shifting to lower energy and vanishes above 12.3 T. 
The energy separation between FP$_{\rm U}$ and FP$_{\rm L}$, $\Delta$$\omega$, increases with increasing field.

\section{Discussion}

Table~\ref{table1} shows the folded phonons which have been found by the FIR studies up to now.  Four folded phonons have been found among the 9 modes predicted by the factor group analysis. 
The reason why the remaining modes are not observed might be due to their weak intensity or concealment in other strong optical phonons. The folded mode of 311.7 cm$^{-1}$ was observed above $H_C$, 
but the details were unclear because it is located on the shoulder of an optical phonon.\cite{Musfeldt} The field dependence of the folded modes of 284.2 and 800 cm$^{-1}$ have not been reported yet.

\begin{figure}
\centerline{\includegraphics*[width=8cm]{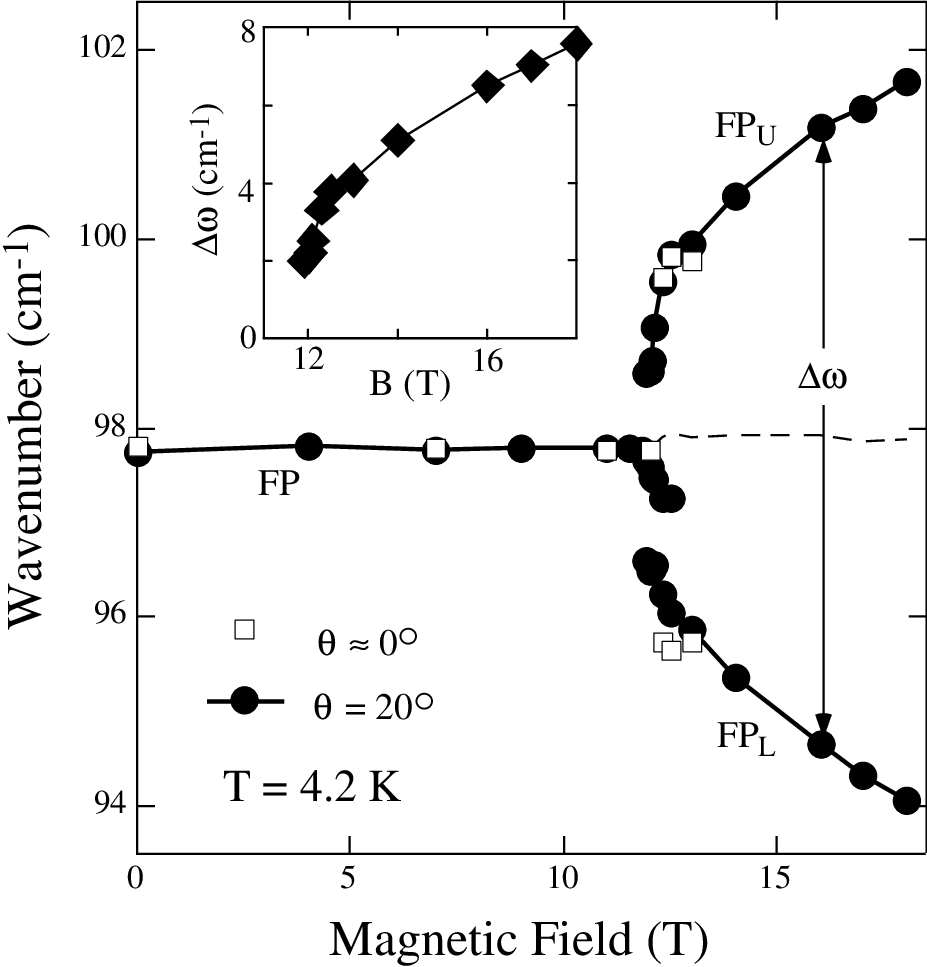}}
\caption{Field dependence of the peak positions of the absorption at 98 cm$^{-1}$ and its satellites at 4.2 K, when the sample is rotated by a few degrees (open squares) and 20$^{\circ}$ (closed circles). 
The dashed line indicates the field dependence of the average energy of the satellites. Inset shows the field dependence of the energy separation between these satellites, $\Delta$$\omega$.}
\label{fig7}
\end{figure}

\begin{figure}
\centerline{\includegraphics*[width=8cm]{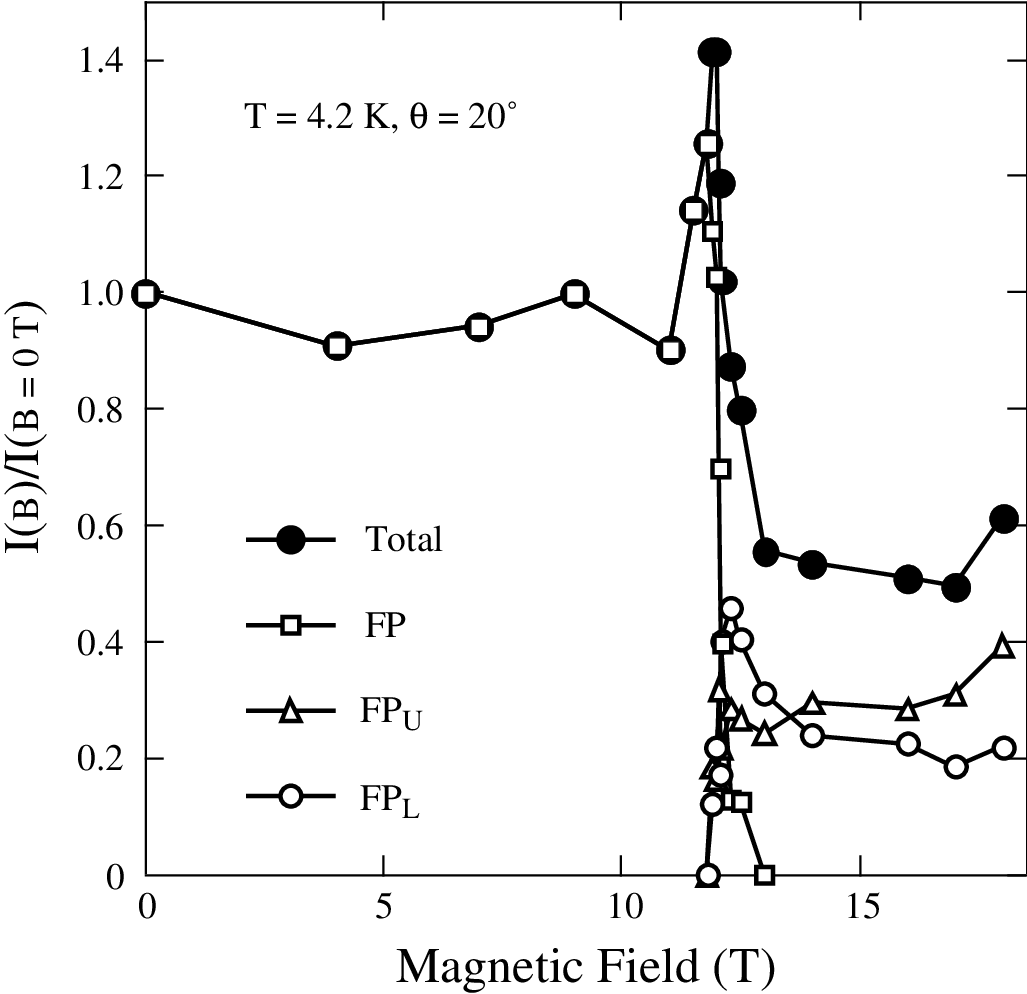}}
\caption{Field dependence of the integrated intensities of the absorption at 98 cm$^{-1}$, its satellites and their total at 4.2 K, normalized to zero field.}
\label{fig8}
\end{figure}

The peak positions of FP, FP$_{\rm U}$ and FP$_{\rm L}$ at $\theta$ = 20$^{\circ}$ are shown in the Fig.~\ref{fig7} as functions of magnetic field.  The absorptions, FP$_{\rm U}$ and FP$_{\rm L}$, were confirmed in the spectrum of the IC phase in Fig.~\ref{fig5}, although their intensity is quite small. 
The peak positions at $\theta$ $\approx$ 0$^{\circ}$, which are also plotted in Fig.~\ref{fig7}, are in good agreement with the result at $\theta$ = 20$^{\circ}$, where deviations of $H_C$ caused by the inclination are quite small. 
This agreement indicates that appearance of FP$_{\rm U}$ and FP$_{\rm L}$ instead of FP in the IC phase is not caused by the sample inclination, but is an intrinsic property of this mode. FP is observed in the D phase and both FP$_{\rm U}$ and FP$_{\rm L}$ are observed in the IC phase. 
They coexist in the vicinity of the boundary between the D and IC phases, which is quite similar to the behavior of the superlattice reflection in the x-ray mesurements.\cite{Kiryukhin1} 
The energy of FP is independent of the static magnetic field and the temperature in the D phase. It slightly shifts toward lower energy in the coexistence region, which may be caused by a strong decrease of the spontaneous strain when going from the D to IC phase,\cite{Takehana2} or by discommensurations which appear in the vicinity of the phase boundary. 
The positions of FP$_{\rm U}$ and FP$_{\rm L}$ are almost symmetrical with respect to that of FP, which indicates that FP line splits into two lines; FP$_{\rm U}$ and FP$_{\rm L}$, in the IC phase. 
$\Delta$$\omega$ increases steeply with increasing field at the vicinity of the phase boundary and the rate of the increase slows down gradually with increasing field (see inset of Fig.~\ref{fig7}). The field dependence of the integrated intensities of FP, FP$_{\rm U}$ and FP$_{\rm L}$ and their total are displayed in Fig.~\ref{fig8}. 
The intensity of FP decreases steeply to zero just above $H_C$ after enhancement at the vicinity of the phase boundary, while it is almost field independent below 10 T. 
Similar enhancements around $H_C$ were observed in the field dependence of both FP$_{\rm U}$ and FP$_{\rm L}$.  Total intensity in the IC phase decreases to about 50$\%$ of zero-field value and gradually decreases with increasing field. 
This behavior is quite similar to that of other folded phonons except for the enhancement around $H_C$.\cite{Loosdrecht2,Loa2} The enhancement was also observed in the Fano type mode at 107 cm$^{-1}$, but was much more moderate.\cite{Loa2} 
The intensity of folded phonons is closely related to the lattice distortion induced by the SP transition, as well as to the intensity of the superlattice reflections. For this reason, the field dependence of the intensity of the folded phonon,\cite{Loa2} the superlattice reflection\cite{Kiryukhin1} and the spontaneous strain\cite{Takehana2} quite resemble each other. 
However, enhancements around $H_C$ were only observed in this mode and a Fano type folded mode in the Raman experiments, which would suggest the presence of some interaction which causes the splitting of this mode in the IC phase. 

\begin{figure}
\centerline{\includegraphics*[width=8cm]{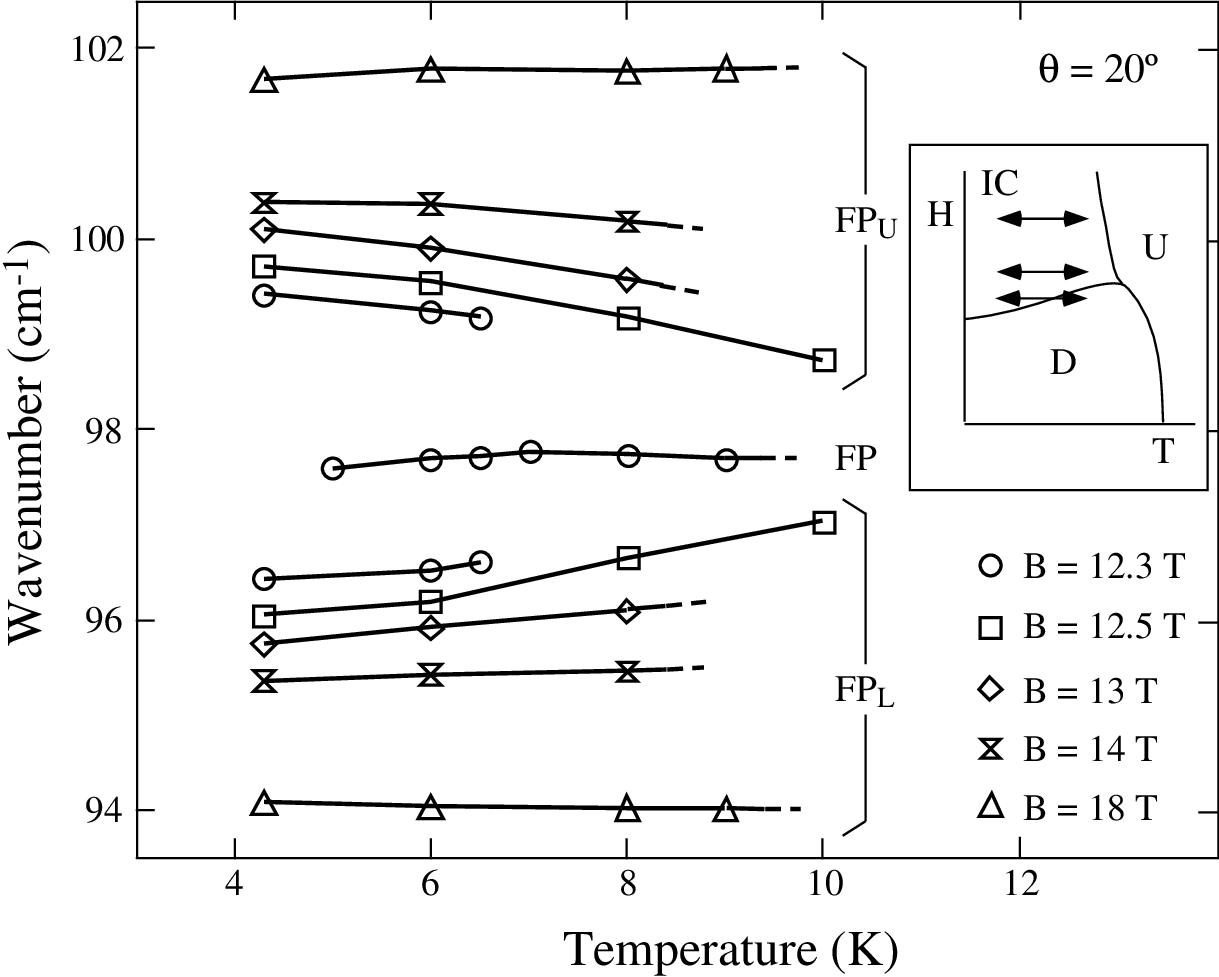}}
\caption{Temperature dependence of the peak positions of FP, FP$_{\rm U}$ and FP$_{\rm L}$ at various fixed fields. The inset shows the schematic phase diagram in the ($H, T$) plane.}
\label{fig9}
\end{figure}

The temperature dependence of the peak positions of FP, FP$_{\rm U}$ and FP$_{\rm L}$ in various fixed fields is shown in Fig.~\ref{fig9}.  
The symmetrical location of FP$_{\rm U}$ and FP$_{\rm L}$ with respect to FP is also confirmed in Fig.~\ref{fig9}. 
The intensity decreases steeply with approaching the D-U phase or the IC-U phase boundary, and these peaks disappear in the U phase. 
FP$_{\rm U}$ and FP$_{\rm L}$ shift toward lower and higher energy, respectively, which means $\Delta$$\omega$ decreases, with increasing temperature. 
The temperature dependence of the peak positions becomes weak with increasing field, and the peak positions at 18 T are almost independent of temperature. 
A possible explanation for this temperature dependence is as follows: $\Delta$$\omega$ decreases with approaching the D-IC phase boundary, corresponding to the field dependence of $\Delta$$\omega$ as shown in Fig.~\ref{fig7}. 
When the magnetic field is just above $H_C$, the magnetic state approaches the phase boundary with increasing temperature and this tendency gets weaker for $H \gg H_C$, according to the shape of the phase boundary (see the inset of Fig.~\ref{fig9}). 

\begin{figure}
\centerline{\includegraphics*[width=8cm]{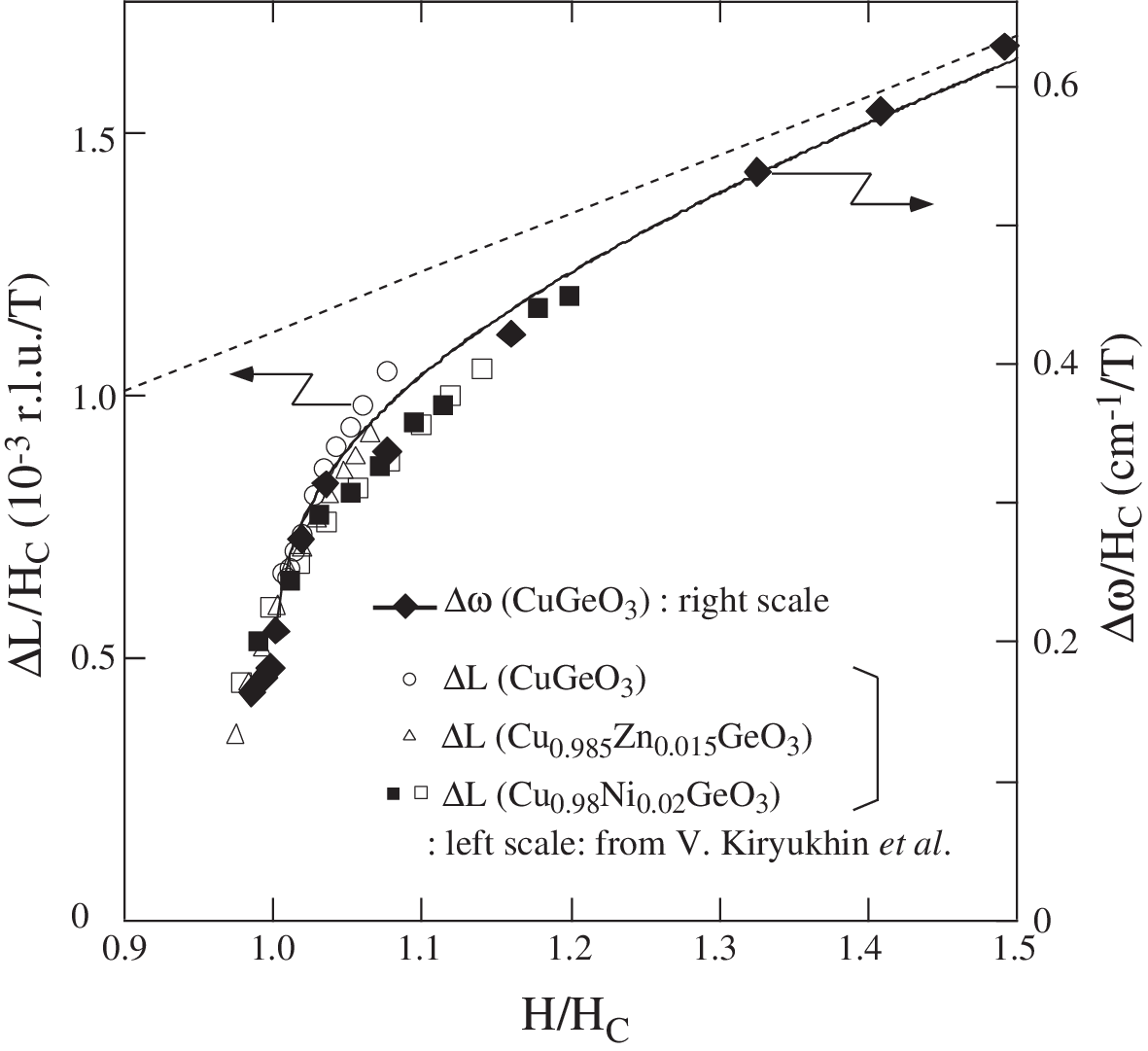}}
\caption{
Field dependence of the energy separation between the satellites, $\Delta$$\omega$, and that of the incommensurability,  $\Delta$L, which were measured by Kiryukhin {\it et al.} (Ref.~\protect\onlinecite{Kiryukhin2}). 
Both $\Delta$$\omega$ and $\Delta$L are scaled by the respective critical field, $H_C$, and can be well described by the theoretical curve $1/\ln[8H_C/(H-H_C)]$ within the mean field theory (Ref.~\protect\onlinecite{Buzdin}). The dashed line indicates the theoretical prediction by Cross for the high magnetic field limit (Ref.~\protect\onlinecite{Cross}).}
\label{fig10}
\end{figure}

To our knowledge, this is the first example of the observation of splitting of the folded phonon in the IC phase. The field dependence of $\Delta$$\omega$ quite resembles that of the incommensurability, $\Delta$L, which was estimated by the splitting of the (3.5 1 2.5) superlattice reflection in the x-ray diffraction experiments.\cite{Kiryukhin2} 
The temperature dependence of $\Delta$$\omega$ is also consistent with that of $\Delta$L, while the latter was investigated only just above $H_C$.\cite{Kiryukhin1} 
We compared the field dependence of $\Delta$$\omega$ and $\Delta$L by scaling them with $H_C$, as shown in Fig.~\ref{fig10}. Critical behavior of $\Delta$L in pure and diluted CuGeO$_{3}$ can be scaled by a universal curve, although there are some deviations dependent on the composition. The universal curve is well described by the function $1/\ln[8H_C/(H-H_C)]$, which is predicted by the mean field theory.\cite{Buzdin}
The right scale was adjusted to fit $\Delta$$\omega$ to the universal curve. It is obvious that $\Delta$$\omega$ is also well described by the same universal curve. 
Moreover, $\Delta$$\omega$ approaches gradually the theoretically predicted curve in the high magnetic field limit by Cross.\cite{Cross} These results suggest that $\Delta$$\omega$ is proportional to the incommensurability, at least up to 18 T. 

In principle, optical excitations are allowed to be infrared active only at ${\bf k}$ = 0 owing to the momentum conservation rule. When the lattice modulation is induced, excitations at ${\bf k}$ = $\pm$ $n{\bf q}$ are allowed in addition to those at ${\bf k}$ = 0, where ${\bf q}$ is the modulation wave vector and $n$ is an integer, although the intensities decrease rapidly with increasing $n$.\cite{Janssen} 
In the case of the SP transition, ${\bf q}$ = ${\bf q}_{SP}$, where ${\bf q}_{SP}$ is the modulation wave vector caused by the SP phase transition, which is located on the zone boundary in the U phase. 
This means that the excitations at the zone boundary in the U phase are folded to the zone center. 
This makes the folded phonons observable below $T_{SP}$. 
In the IC phase, ${\bf q}$ deviates from ${\bf q}_{SP}$ by a certain wave number $\Delta$${\bf q}$, which makes the excitations at ${\bf k}$ = $\pm({\bf q}_{SP} - \Delta{\bf q})$ infrared active. 
These modes have the same wavelength as the incommensurability and propagate along the $c$ axis toward the opposite direction from each other. 
Experimentally, the average component of the spin polarization was confirmed to have a periodicity of 2$\Delta{\bf q}$ by the NMR\cite{Horvatic} and the neutron scattering studies\cite{Ronnow}, while the staggered component has a periodicity of $\pm({\bf q}_{SP} - \Delta{\bf q})$. 
Note that 2$\Delta{\bf q}$ is equivalent with the difference between $\pm({\bf q}_{SP} - \Delta{\bf q})$. 
When the two modes of ${\bf k}$ = $\pm({\bf q}_{SP} - \Delta{\bf q})$ have the strong spin-phonon coupling, mixing of these modes and the splitting of the mixed modes are caused by the interaction between these two modes through the modulation of the spin polarization with the periodicity of 2$\Delta{\bf q}$. 
Therefore, if we define $\kappa$ as the coupling between the $\pm({\bf q}_{SP} - \Delta{\bf q})$ modes, $\kappa$ is certainly nonzero. 
The energy separation can be described as follows:

\begin{equation}
\left|
\begin{array}{cc}
\omega_0^2-\omega^2 & \kappa\\
\kappa* & \omega_0^2-\omega^2
\end{array}
\right| = 0,
\end{equation}

\begin{equation}
\omega^2 = \omega_0^2 \pm |\kappa|,
\end{equation}

\begin{equation}
\omega \approx \omega_0 \pm \frac{|\kappa|}{2\omega_0},\label{equ3}
\end{equation}
where $\omega$ is the absorption energy with the coupling, $\omega$$_{0}$ is that without the coupling. 
Equation~(\ref{equ3}) means that the absorption is split into two components and they appear on both sides of $\omega$$_{0}$ with the same separations, which is quite consistent with the experimental results. 
It also suggests that $\Delta$$\omega$ is proportional to $|\kappa|$. 
The eigenstates with the energy $\omega_0 \pm \frac{|\kappa|}{2\omega_0}$ are two kinds of the standing waves having the same period as the incommensurability and different phases. 
$|\kappa|$ is dependent on both temperature and magnetic field, because the incommensurability are strongly dependent on them. 
The field dependence of $\Delta\omega$ is in good agreement with that of the magnetization.\cite{Horvatic} 
The temperature dependence of $\Delta\omega$ is also consistent with that of the magnetization.\cite{Hori} 
It would be a reasonable assumption that $|\kappa|$ is proportional to the magnetization, i.e., the density of the solitons, because we argued that the coupling to the average component of the spin polarization causes the energy splitting. 
Therefore, $|\kappa|$ is proportional to $\Delta$$q$, which leads to the relation that $\Delta$$\omega$ is proportional to $\Delta$$q$. 
This conclusion is quite consistent with our experimental results. 
Note that there is no other example than FP to have the splitting of the folded mode in the IC phase, while the above argument may be applied also to any other modes. 
This indicates that FP is the particular mode to have the strong coupling to the average component of the spin polarization. 
In general, the spin-Peierls system has the spin-phonon coupled modes, and one of them is the soft mode that softens toward $T_{SP}$.\cite{Cross} 
The soft phonon mode has not been found yet, and is now believed to be nonexistent in CuGeO$_3$. 
Since CuGeO$_3$ is a spin-Peierls system, there should be the spin-phonon coupled modes, even though there is no soft mode. 
Braden {\em et al.} argued the possibility of the spin-phonon coupling of the phonon modes at the zone boundary which become the Raman active modes in the D phase, and observed the anomalous hardening of these two modes with lower frequencies above $T_{SP}$ in the INS experiments, while no soft mode behavior was observed around $T_{SP}$.\cite{Braden2} 
The anomalous hardening was explained as the pretransitional fluctuations of the Peierls-active phonon modes, by reconsidering the Cross and Fischer's approach.\cite{Gros} 
The energy of the FP mode is the lowest among those of the folded phonon modes which have been found up to now, and is just close to that of the lowest folded mode with Raman activity, which was argued to have the strong spin-phonon coupling. 
The splitting in the IC phase is considered as the proof that FP has the strong spin-phonon coupling. 
It also indicates the possibility of FP to be one of the phonon modes contributing to the mechanism of the spin-Peierls transition, similar to the Raman active modes as was argued in Ref.~\onlinecite{Braden2}. 
The intensities of both of the split-off branches of the superlattice reflection in the IC phase would be a quarter of that in the D phase, on the assumption that the lattice modulation in the IC phase is described as the sinusoidal wave $\delta\cos(\Delta{\bf q}\cdot{\bf r})$, where $\delta$ is the magnitude of the lattice modulation in the D phase. 
This was confirmed in the experimental results.\cite{Kiryukhin1} 
Similarly, the intensities of the folded phonon modes at ${\bf k}$ = $\pm({\bf q}_{SP} - \Delta{\bf q})$ in the IC phase would be a quarter of that at ${\bf k}$ = ${\bf q}_{SP}$ in the D phase; that is, the total intensity in the IC phase is a half in the D phase. 
This is roughly in agreement with the results of this study. 
The intensities of other folded modes just above $H_C$ also decrease to about half of that in the D phase.\cite{Loa2} 
The enhancement of the intensity around $H_C$ might be caused by the critical fluctuation of the spin system, because no anomaly was observed in the x-ray measurements. 

Next we will discuss the new broad absorption band M2. Table~\ref{table2} shows the magnetic excitations which have been found by the FIR experiments. The absorption of 19 cm$^{-1}$ has been observed only by using the FIR laser.\cite{Nojiri} 
M2 was found for the first time in this study, and the shift under the magnetic field indicates that M2 has a magnetic origin. 
Absorptions of both 19 and 44 cm$^{-1}$ were assigned to the magnetic excitation from the singlet ground state to the triplet states, which means that there are two magnetic gaps at the $\Gamma$ point. 
A second magnetic excitation branch, whose energy gap corresponds to 44 cm$^{-1}$ at the $\Gamma$ point, was found by the INS experiments.\cite{Lorenzo2} 
The newly observed branch was assigned to the zone-folded one due to the existence of the DM antisymmetric exchange terms, which means that the energy gap of the new branch at the $\Gamma$ point corresponds to that of the original branch at the zone boundary. 
The energy scan spectrum for the zone boundary {\bf Q} = (0, 2, 0.5) was investigated by the INS experiments (See Fig.~2 of Ref.~\onlinecite{Ain}), and a continuum of magnetic excitations, whose peak is located at 7 meV, was found to be separated from the triplet branch by the second gap of 2 meV. 
We infer that M2 originates in a magnetic excitation at the $\Gamma$ point from the singlet ground state to the continuum state spreading over higher energy region above the newly observed triplet branch. 
Note that not only the energy of M1, but also that of M2 coincide with those of the first and second peaks at 5 meV and 7 meV, respectively, on the data for {\bf Q} = (0, 2, 0.5) in the INS experiments (See Fig.~2 of Ref.~\onlinecite{Ain}). 
The line shapes of M1 and M2 also resemble those of the corresponding ones in the INS experiments. 
There are other similar features such as the ratio of two peaks and the long tail on the higher energy side of the continuum of magnetic excitations. 
An ^^ ^^ absorption valley" between M1 and M2 corresponds to the ^^ ^^ second gap" found by the INS measurements. 
A possibility of an optical excitation between the singlet ground state and the continuum state was suggested by Kokado and Suzuki.\cite{Kokado} 
The magnetic excitation at 19 cm$^{-1}$ was found to be too weak to be observed in our experiments. 
Missing of the absorption at 19 cm$^{-1}$ in this study would be connected to the small density of state of the magnetic excitation branches. 
The numerical studies without the DM interaction predicted that the density of states at the zone center is much smaller than at the zone boundary.\cite{Haas} 
The ^^ ^^ rampart" structure of the continuum state, which was found in the INS experiments around 30 meV,\cite{Arai} is expected to be observed, but has not been confirmed in our spectrum probably because of the strong absorption of other origins, as shown in Fig.~\ref{fig1}.
For the first time, we have not only observed the magnetic excitation from the singlet state to the continuum state by the optical method, but also confirmed the presence of the second gap located at 5-7 meV at the $\Gamma$ point, which have not yet been observed in the INS experiments.

\section{Conclusions}

We have performed the polarized FIR spectroscopic measurements and the unpolarized FIR magneto-optical studies on the spin-Peierls compound CuGeO$_{3}$. A sharp absorption line, which appears at 98 cm$^{-1}$ in the D phase, was assigned to a folded phonon mode of B$_{3u}$ symmetry. 
This folded mode was found to split into two components in the IC phase. 
The energy separation of the two split branches is proportional to the incommensurability in the IC phase. 
A new broad absorption centered at 63 cm$^{-1}$ was found in the ${\bf E}\parallel b$ axis polarization spectra, which was assigned to magnetic excitation from singlet ground state to a continuum state.

\begin{table}
\caption{Folded phonon modes which appear in the FIR spectra below $T_{SP}$. The first one is the new mode found in this study.}
\label{table1}
\begin{tabular}{ccc}
Frequency $\omega$ (cm$^{-1}$) & Polarization & IC hpase (above $H_C$)\\
\tableline
98 & ${\bf E}\parallel a$ & splitting\\
284.2\tablenote{from Ref.\onlinecite{Popova}} & ${\bf E}\parallel c$ & ---\\
311.7\tablenote{from Ref.~\onlinecite{Popova}} & ${\bf E}\parallel b$ & detectable\tablenote{from Ref.~\onlinecite{Musfeldt}}\\
800\tablenote{from Ref.\onlinecite{Damascelli}} & ${\bf E}\parallel b$ & ---\\
\end{tabular}
\end{table}

\begin{table}
\caption{Magnetic absorptions which appear in the FIR spectra below $T_{SP}$. The last one is the new absorption band found in this study.}
\label{table2}
\begin{tabular}{cccc}
Frequency $\omega$ (cm$^{-1}$) & Polarization & D phase & IC phase\\
(at 4.2 K) & (at 0 T) & (below $H_C$) & (above $H_C$)\\
\tableline
19\tablenote{from Ref.~\onlinecite{Nojiri}} & --- & splitting & ---\\
44\tablenote{from Ref.~\onlinecite{Brill}} & ${\bf E}\parallel b > {\bf E}\parallel c$ & splitting & ---\\
63 & ${\bf E}\parallel b > {\bf E}\parallel c$ & shifting & broadening\\
\end{tabular}
\end{table}
\end{multicols}

\begin{references}
\bibitem{Hase} M. Hase, I. Terasaki, and K. Uchinokura, Phys. Rev. Lett. {\bf 70}, 3651 (1993).
\bibitem{Kaminura} O. Kamimura, M. Terauchi, M. Tanaka, O. Fujita, and J. Akimitsu, J. Phys. Soc. Jpn. {\bf 63}, 2467 (1994).
\bibitem{Pouget} J. P. Pouget, L. P. Regnault, M. Ain, B. Hennion, J. P. Renard, P. Veillet, G. Dhalenne, and A. Revcolevschi, Phys. Rev. Lett. {\bf 72}, 4037 (1994).
\bibitem{Hitota} K. Hirota, D. E. Cox, J. E. Lorenzo, G. Shirane, J. M. Tranquada, M. Hase, K. Uchinokura, H. Kojima, Y. Shibuya, and I. Tanaka, Phys. Rev. Lett. {\bf 73}, 736 (1994).
\bibitem{Hase2} M. Hase, I. Terasaki, K. Uchinokura, M. Tokunaga, N. Miura, and H. Obara, Phys. Rev. B {\bf 48}, 9616 (1993).
\bibitem{Kiryukhin1} V. Kiryukhin and B. Keimer, Phys. Rev. B {\bf 52}, R704 (1995).
\bibitem{Kiryukhin2}  V. Kiryukhin, B. Keimer, J.P. Hill, and A. Vigliante, Phys. Rev. Lett. {\bf 76}, 4608 (1996).
\bibitem{Lorenzo} J. E. Lorenzo, K. Hirota, G. Shirane, J. M. Tranquada, M. Hase, K. Uchinokura, H. Kojima, I. Tanaka, and Y. Shibuya, Phys. Rev. B {\bf 50}, 1278 (1994).
\bibitem{Takehana1} K. Takehana, M. Oshikiri, G. Kido, M. Hase, and K. Uchinokura, J. Phys. Soc. Jpn. {\bf 65}, 2783 (1996).
\bibitem{Takehana2} K. Takehana, T. Takamasu, M. Hase, G. Kido, and K. Uchinokura, Physica B {\bf 246-247}, 246 (1998).
\bibitem{Vollenkle} H. V{\" o}llenkle, A. Wittmann, and H. Nowotny, Monatsh. Chem. {\bf 98}, 1352 (1967).
\bibitem{Popovic} Z. V. Popovi{\' c}, S. D. Devi{\' c}, V. N. Popov, G. Dhalenne, and A. Revcolevschi, Phys. Rev. B {\bf 52}, 4185 (1995).
\bibitem{Braden} M. Braden, G. Wilkendorf, J. Lorenzana, M. A{\" i}n, G. J. McIntyre, M. Behruzi, G. Heger, G. Dhalenne, and A. Revcolevschi, Phys. Rev. B {\bf 54}, 1105 (1996).
\bibitem{Ogita} N. Ogita, T. Minami, Y. Tanimoto, O. Fujita, J. Akimitsu, P. Lemmens, G. G{\" u}ntherodt, and M. Udagawa, J. Phys. Soc. Jpn. {\bf 65}, 3754 (1996).
\bibitem{Loa} I. Loa, S. Gronemeyer, C. Thomsen, and R. K. Kremer, Solid State Commun. {\bf 99}, 231 (1996).
\bibitem{Loosdrecht} P. H. M. van Loosdrecht, J. P. Boucher, G. Martinez, G. Dhalenne, and A. Revcolevschi, Phys. Rev. Lett. {\bf 76}, 311 (1996).
\bibitem{Damascelli} A. Damascelli, D. van der Marel, F. Parmigiani, G. Dhalenne, and A. Revcolevschi, Phys. Rev. B {\bf 56}, R11373 (1997).
\bibitem{Popova} M. N. Popova, A. B. Sushkov, S.A. Golubchik, A.N. Vasil'ev, and L. I. Leonyuk, Phys. Rev. B {\bf 57}, 5040 (1998).
\bibitem{Takehana3} K. Takehana, M. Oshikiri, T. Takamasu, M. Hase, G. Kido, and K. Uchinokura, J. Magn. Magn. Mater. {\bf 177-181}, 699 (1998).
\bibitem{Loosdrecht2} P. H. M. van Loosdrecht, J. P. Boucher, G. Martinez, G. Dhalenne, and A. Revcolevschi, J. Appl. Phys. {\bf 79}, 5395 (1996).
\bibitem{Loa2} I. Loa, S. Gronemeyer, C. Thomsen, and R. K. Kremer, Z. Phys. Chem. {\bf 201}, 333 (1997).
\bibitem{Brill} T. M. Brill, J. P. Boucher, J. Voiron, G. Dhalenne, A. Revcolevschi, and J. P. Renard, Phys. Rev. Lett. {\bf 73}, 1545 (1994).
\bibitem{Li} G. Li, J. L. Musfeldt, Y. J. Wang, S. Jandl, M. Poirier, A. Revcolevschi, and G. Dhalenne, Phys. Rev. B {\bf 54}, R15633 (1996).
\bibitem{Loosdrecht3} P. H. M. van Loosdrecht, S. Huant, G. Martinez, G. Dhalenne, and A. Revcolevschi, Phys. Rev. B {\bf 54}, R3730 (1996).
\bibitem{Nojiri} H. Nojiri, H. Ohta, N. Miura, and M. Motokawa, Physica B {\bf 246-247}, 16 (1998).
\bibitem{Kuroe} H. Kuroe, T. Sekine, M. Hase, Y. Sasago, K. Uchinokura, H. Kojima, I. Tanaka, and Y. Shibuya, Phys. Rev. B {\bf 50}, 16468 (1994).
\bibitem{Lorenzo2} J. E. Lorenzo, L. P. Regnault, J. P. Boucher, B. Hennion, G. Dhalenne, and A. Revcolevschi, Europhys. Lett. {\bf 45}, 619 (1999).
\bibitem{Yamada} I. Yamada, M. Nishi, and J. Akimitsu, J. Phys.: Condens. Matter {\bf 8}, 2625 (1996).
\bibitem{Uhrig} G. S. Uhrig, Phys. Rev. Lett. {\bf 79}, 163 (1997).
\bibitem{Takehana4} K. Takehana, M. Oshikiri, G. Kido, A. Takazawa, M. Sato, K. Nagasaka, M. Hase, and K. Uchinokura, Physica B {\bf 216}, 354 (1996).
\bibitem{Harris} Q. J. Harris, Q. Feng, R. J. Birgeneau, K. Hirota, K. Kakurai, J. E. Lorenzo, G. Shirane, M. Hase, K. Uchinokura, H. Kojima, I. Tanaka, Y. Shibuya, Phys. Rev. B {\bf 50}, 12606 (1994).
\bibitem{Musfeldt} J. L. Musfeldt, Y. J. Wang, S. Jandl, M. Poirier, A. Revcolevschi, and G. Dhalenne: Phys. Rev. B {\bf 54}, 469 (1996).
\bibitem{Buzdin} A. I. Buzdin, M. L. Kulic, and V. V. Tugushev, Solid State Commun. {\bf 48}, 483 (1983).
\bibitem{Cross} M. C. Cross, Phys. Rev. B {\bf 20}, 4606 (1979).
\bibitem{Janssen} T. Janssen, J. Phys. C {\bf 12}, 5391 (1979).
\bibitem{Horvatic} M. Horvati{\' c}, Y. Fagot-Revurat, C. Berthier, G. Dhalenne, and A. Revcolevschi, Phys. Rev. Lett. {\bf 83}, 420 (1999).
\bibitem{Ronnow} H. M. R$\o$nnow, M. Enderle, D. F. McMorrow, L.-P. Regnault, G. Dhalenne, A. Revcolevschi, A. Hoser, K. Prokes, P. Vorderwisch, and H. Schneider, cond-mat/9912251.
\bibitem{Hori} H. Hori, M. Furusawa, T. Takeuchi, S. Sugai, K. Kindo, and A. Yamagishi, J. Phys. Soc. Jpn {\bf 63}, 18 (1994).
\bibitem{Braden2} M. Braden, B. Hennion, W. Reichardt, G. Dhalenne, and A. Revcolevschi, Phys. Rev. Lett. {\bf 80}, 3634 (1998).
\bibitem{Gros} C. Gros, and R. Werner, Phys. Rev. B {\bf 58}, R14677 (1998).
\bibitem{Ain} M. A{\" i}n, J. E. Lorenzo, L. P. Regnault, G. Dhalenne, A. Revcolevschi, B. Hennion, and Th. Jolicoeur, Phys. Rev. Lett. {\bf 78}, 1560 (1997).
\bibitem{Kokado} S. Kokado and N. Suzuki, Proc. 4th Int. Symp. on Advanced Physical Fields, Tsukuba, Japan, 1999, p. 243.
\bibitem{Haas} S. Haas and E. Dagotto, Phys. Rev. B {\bf 52}, R14396 (1995).
\bibitem{Arai} M. Arai, M. Fujita, M. Motokawa, J. Akimitsu, and S. M. Bennington, Phys. Rev. Lett. {\bf 77}, 3649 (1996).
\end{references}
\end{document}